# Power Handling Improvement in Cross-Sectional Lame Mode Resonators Operating in the Ku-band´

Luca Spagnuolo[1], Gabriel Giribaldi[1], Filippo Perli[2], Alberto Corigliano[2], Luca Colombo[1], and Matteo Rinaldi[1]
[1]The Institute for NanoSystems Innovation, Northeastern University, Boston MA, USA
[2]Politecnico di Milano, Milan, Italy

*Abstract*—This study presents power handling improvements in cross-sectional Lame-Mode Resonators (CLMRs) designed for´ operation in the Ku-band. Previously fabricated CLMR devices failed at approximately 8 dBm of input power, primarily due to electromigration in the aluminum interdigitated electrodes (IDTs). To better understand this mechanism in CLMRs, a datadriven thermal model is developed to analyze localized heating effects within the resonator body, which are known to accelerate electromigration. Based on insights from this model, Aluminum Silicon Copper (AlSiCu) was selected for the IDTs due to its superior thermal stability and resistance to electromigration.

Devices fabricated with AlSiCu exhibited no signs of performance degradation, with the best-performing resonator achieving a mechanical quality factor ($Q_m$) of 360, a maximum Bode quality factor ($Q_{Bode}$) of 500, and an electromechanical coupling coefficient ($k_t^2$) of 6.3%. Moreover, the use of AlSiCu significantly increased the maximum input power the device can withstand, showing an improvement of up to 6 dBm over previous devices. These improvements in power handling make the devices strong candidates for high-power Ku-band filtering applications.

## I. Introduction

Microacoustic piezoelectric resonators have been pivotal components in modern Radio Frequency (RF) communication paradigms [1]. However, ongoing research continues to expand the potential of micro- and nanoelectromechanical systems (M/NEMS), driving advancements in next-generation applications. In particular, the 5G FR-2 (mmWave) spectrum remains a primary focus, while the Ku-band (spanning 12 to 18 GHz) has gained significant attention due to its relevance for dense urban area coverage, as outlined in the proposed 5G FR-3 framework [2]. At the same time, the increasing complexity of modern wireless systems has underscored the critical role of the RF Front-End (RFFE) in ensuring optimal performance [3]. These modules must support high-output power levels to enable efficient transmission, particularly in the 5G and emerging FR-3 bands where higher bandwidth and higher frequencies are required. In this context, high power handling is fundamental to meeting the demands of advanced wireless applications. This is particularly true for applications like satellite communications (SATCOM) [4] [5] and mobile devices operating in high-frequency bands, where higher transmission power is required to overcome propagation losses and interference.

The development of advanced piezoelectric materials such as Scandium-doped Aluminum Nitride (ScAlN) [6] [7] [8], and Scandium-Boron co-doped Aluminum Nitride (AlScBN) [9] has played a pivotal role in addressing the challenges posed by high-frequency and high-power RF applications [10] [11] [12]. Among resonator technologies that can exploit ScAlN, cross-sectional Lame-Mode Resonators (CLMRs) [13] are of´ particular interest due to their frequency lithographic tunability [14]. However, previously fabricated CLMRs [15] were limited to a maximum input power of approximately 8 to 9 dBm before failure.

In this work, we investigate power handling in crosssectional Lame-mode resonators (CLMRs) operating in the´ Ku-band. A data-driven thermal model is developed to analyze localized heating effects and understand the mechanisms leading to failure. Based on these insights, Aluminum Silicon Copper is used to replace pure Aluminum in the interdigitated transducers (IDTs), improving thermal stability and resistance to electromigration. This work presents newly fabricated CLMRs capable of handling input power levels exceeding 14 dBm. These devices leverage electrode optimization to enhance power handling while maintaining key performance indicators (KPIs) such as high quality factor (Q) and electromechanical coupling coefficient ($k_t^2$).

## II. Methods

Cross-Sectional Lame Mode Resonators (CLMRs) leverage´ both the $d_{33}$ and $d_{31}$ piezoelectric coefficients to excite a twodimensional acoustic standing wave, which exhibits high stress and displacement components in the lateral (*x*) and thickness (*z*) directions (Fig. 1a).

A critical design parameter in CLMRs is the interdigitated electrode configuration, where both the material selection and geometric dimensions directly influence the resonant frequency, coupling, and quality factor [16]. While aluminum has been commonly employed in the fabrication of CLMRs, its poor resistance to electromigration has resulted in limited power handling capabilities [15].

Electromigration [17] is a diffusion-driven failure mechanism in microelectronic interconnects, characterized by the transport of metal atoms caused by momentum transfer from high-density electron flow. This process induces atomic migration from one terminal to the other, leading to the formation of voids, that can cause open circuits, and hillocks that can cause short circuits. Atomic diffusion primarily occurs along high-diffusivity paths such as grain boundaries. Moreover, the phenomenon is strongly temperature-dependent and is significantly accelerated by Joule heating ($I^2R$ losses), which raises the local temperature, enhancing diffusion rates leading to a thermal runaway.

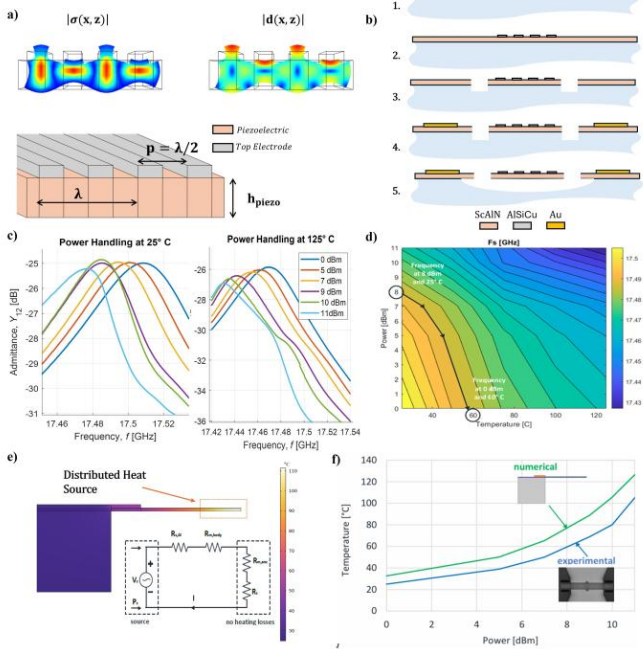

Fig. 1: (a) Cross-sectional Lamé Mode Resoantors with´ Von Mises equivalent stress ans displacement magnitude; (b) Micro-fabrication process flow: Top to bottom: 1) ScAlN reactive sputtering; 2) Top electrode patterning via electron beam (e-beam) lithography, aluminum-silicon-copper (AlSiCu) thermal evaporation, and lift-off; 3) Opening of the etch pits by ion milling; 4) Pads and interconnects patterning via direct laser lithography; 5) $XeF_2$ isotropic etch; (c) Measurement test for the extraction of the thermal properties; (d) Inferred temperature contour plot ; (e) Data-Driven thermal model; (f) Experimental vs. numerical extracted temperature.

*A. Thermal modeling*

To investigate these thermal effects in Cross-Sectional Lamé´ Mode Resonators (CLMRs), a data-driven thermal modeling approach is used. The device is experimentally characterized under varying temperature and RF power conditions, with temperatures ranging from 0°C to 125°C and input powers from 0 dBm to 11 dBm (Fig. 1c).

Measurements are performed using a 2-port configuration with a Vector Network Analyzer (VNA), and the device is mounted on a temperature-controlled heated stage. Resonant frequency shifts are extracted across these conditions, enabling the inference of the device's thermal behavior (Fig. 1d). This empirical data is then used to calibrate and refine a COMSOL™ Multiphysics-based heat transfer model (Fig. 1e), where heating losses are modeled using an equivalent electrical circuit representation.

In this model, the thermal effects are primarily attributed to losses within the resonator body and Joule heating of IDTs. Ohmic losses in the interconnects are neglected, as their contribution to total power dissipation is small compared to losses in the active area of the CLMR. Additionally, anchor losses, which represent mechanical energy dissipation at the support points, are not included in the model since they are negligible in this technology. The temperature response obtained from the COMSOL™ model is compared with the experimental measurements (Fig. 1f). While the two results exhibit similar trends, they do not perfectly overlap. To further improve the accuracy of the model, the measured temperature of the device must be incorporated.

*B. Fabrication*

To mitigate electromigration, a widely used approach is to use Aluminum Silicon Copper (AlSiCu) alloys. The superior resistance of AlSiCu to electromigration is primarily attributed to the grain boundary segregation of copper, which significantly inhibits the diffusion of aluminum atoms across grain boundaries [18]. This segregation effect enhances the structural stability of the IDTs, enabling improved power handling and long-term reliability.

The fabrication process used for the CLMRs is depicted in Fig. 1b. First, 150 nm of 30% ScAlN is deposited in-house on a 8-in high-resistivity silicon substrate through reactive sputtering (Evatec CLUSTERLINE® 200 II system). Then, interdigitated fingers (IDTs) with a width of 100 nm are patterned via e-beam lithography to target the desired frequency of 18 GHz. A 50 nm film of Aluminum Silicon Copper (AlSiCu) is deposited by thermal evaporation, followed by an overnight lift-off in Microposit Remover 1165. Next, etch pits are opened to expose the underlying silicon by ion milling through the ScAlN layer. This is followed by the patterning of contact pads using maskless laser lithography. A 250 nm gold layer is then deposited by electron beam evaporation, followed by an overnight lift-off. Finally, the device is released by isotropic etching of the silicon substrate using $XeF_2$. The final fabricated device is shown in Fig. 2a.

III. RESULTS

*A. Resonator Performance*

The fabricated devices are characterized under standard laboratory conditions (T = 293 K) using 150 $\mu$ ground–signal–ground (GSG) probes in a two-port configuration to evaluate their frequency response. Scattering (S-) parameters are measured using a Keysight P5008A vector network analyzer and subsequently converted to admittance (Y-) parameters via software. The equivalent circuit parameters are extracted by fitting the transmission admittance ($Y_{12}$) response to a modified Butterworth–Van Dyke (mBVD) equivalent circuit model [19].

The admittance response of the device exhibiting the highest 3-dB quality factor ($Q_{3dB}$) is shown in Fig. 2c. This device achieves a $Q_{3dB}$ of 251, a motional quality factor ($Q_m$) of 360, and an electromechanical coupling coefficient ($k_t^2$) of 6.3%, resulting in an overall figure of merit ($FOM \approx Q_m \cdot k_t^2$) exceeding 22. Additionally, a Bode quality factor ($Q_{Bode}$) of 500 is extracted from the same device, as illustrated in Fig. 2b. To assess the impact of AlSiCu fingers on resonator performance, a comparison table with devices at a similar operating frequency is reported (Table I).

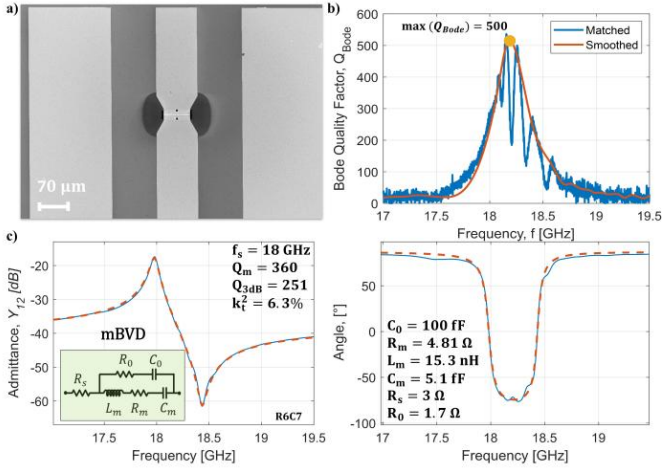

Fig. 2: (a) SEM picture of final device; (b) $Q_{Bode}$ profile and extracted maximum value from smoothing; (c) Admittance response ($Y_{12}$) and phase ($\theta$) of a fabricated CLMR operating at 18 GHz and exhibiting the largest 3-dB quality factor ($Q_{3dB}$).

| Ref | Material | Type | $f_s$ [GHz] | $Q_{3dB}$ | $k_t^2$ [%] |
|---|---|---|---|---|---|
| This Work | ScAlN | CLMR | 18 | 251 | 6.3 |
| [20] | ScAlN | CLMR | 17.76 | 208 | 5.58 |
| [20] | ScAlN | CLMR | 18.83 | 182 | 6.87 |
| [20] | ScAlN | CLMR | 18.17 | 209 | 6.12 |
| [21] | ScAlN | FBAR | 21.4 | 62 | 7 |
| [22] | YZ30° X-Cut LN | d-LVR | 14.47 | 105 | 4 |
| [22] | YZ30° X-Cut LN | d-LVR | 16.21 | 55 | 5.8 |
| [23] | AlN | CLMR | 26.1 | 363 | 1 |
| [24] | Epi-AlN on SiC | FBAR | 14.73 | 155 | 2.3 |

TABLE I

The results indicate that the use of AlSiCu does not introduce any degradation in device performance. On the contrary, the resonators fabricated with AlSiCu exhibit superior figures of merit, achieving the highest reported values for both the 3dB-quality factor ($Q_{3dB}$) and the electromechanical coupling coefficient ($k_t^2$) at 18 GHz.

### B. Aluminum vs Aluminum Silicon Copper

The power limitations in previous designs are primarily attributed to the electromigration of aluminum (Al) IDTs. This phenomenon, driven by high current densities and thermal stress, leads to the formation of gaps in the Al IDTs, ultimately resulting in device failure (Fig. 3b).

By using AlSiCu, the amount of current density that an IDT finger can withstand is significantly increased. To quantify this improvement, a comparison is made between two devices of similar sizes. Fig. 3a presents the results of a power handling test ranging from 0 dBm to 14 dBm. The device with Al IDTs

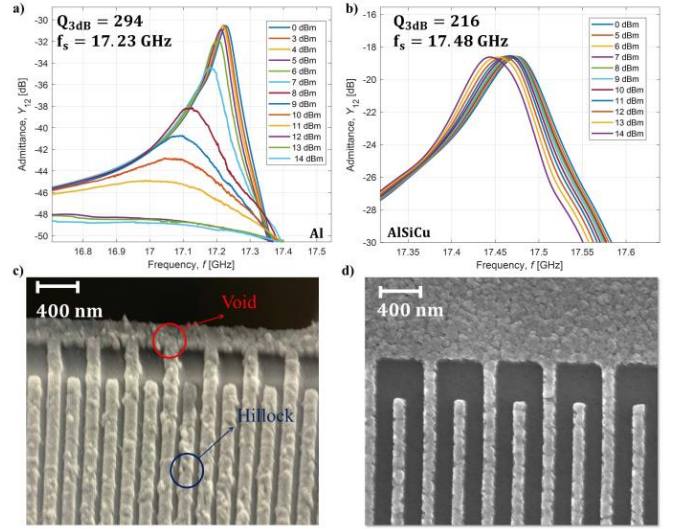

Fig. 3: (a) Power handling test with input power from 0 dBm to 14 dBm; (b) SEM of Aluminum fingers after failure; (c) SEM of Aluminum Silicon Copper fingers

(left) begins to exhibit a degraded response at approximately 7 dBm, which ultimately fails at 8-9 dBm. In contrast, the device with AlSiCu IDTs (right) shows no signs of degradation even at 14 dBm.

To further investigate the underlying mechanisms, the current density and the temperature increase of an IDT finger are calculated. First, the total current is extracted from the input voltage $V_{in}$, which corresponds to the voltage delivered by the Vector Network Analyzer (VNA) to a matched 50Ω load.

$$I_{tot} = V_{in}/R_{tot} \quad (1)$$

Here, $R_{tot}$ is the total resistance of the systems which includes the motional resistance of the resonator, the pad resistance, and the source and load impedance of the VNA Then, considering the resistance of each finger in parallel, the current density for a single finger is extracted as in Eq. 2.

$$J_f = 2 \cdot \frac{I_{tot}}{(N_f(W \cdot t_m))} \quad (2)$$

The aluminum finger failed at a current density of $J_{8dBm} = 6.36 \times 10^7 A/cm^2$ which exceeds the known electromigration threshold for aluminum. In contrast, the AlSiCu finger sustained a current density of $J_{14dBm} = 1.8 \times 10^8 A/cm^2$, which is near the known electromigration limit of AlSiCu. However, further increases in input power were not possible due to limitations of the experimental setup.

Notably, the AlSiCu fingers sustained nearly three times the current density tolerated by the aluminum fingers, demonstrating their superior resistance to electromigration.

## IV. Conclusions

In this study, the fabrication of Cross-sectional Lamé Mode Resonators (CLRMs) with enhanced power handling capabilities for Ku-band applications is successfully demonstrated. By addressing the issue of electromigration in aluminum IDTs with aluminum silicon copper alloy, the devices achieved a significant improvement in powers handling performance, withstanding input power levels up to 14 dBm without observable degradation. Furthermore, a data-driven thermal modeling approach is employed to investigate the temperature rise in the resonator under varying input power levels, providing key insights to study the influence of temperature on electromigration. These findings position CLMRs as strong candidates for next-generation high-frequency filtering components in RF front-end modules, particularly in demanding Ku-band applications.

## Acknowledgment

This work was supported by the Defense Advanced Research Project Agency (DARPA) COFFEE project under Contract HR001122C0088 and developed in collaboration with Raytheon Missiles & Defense (RTX) and the U.S. Naval Research Lab (NRL). The authors would also like to thank Northeastern University Kostas Cleanroom and Harvard CNS staff.

## References

[1] A. Hagelauer, R. Ruby, S. Inoue, V. Plessky, K.-Y. Hashimoto, R. Nakagawa, J. Verdu, P. d. Paco, A. Mortazawi, G. Piazza, Z. Schaffer, E. T.T. Yen, T. Forster, and A. Tag, "From Microwave Acoustic Filters to Millimeter-Wave Operation and New Applications," *IEEE Journal of Microwaves*, vol. 3, pp. 484–508, Jan. 2023.

[2] R. Vetury, A. Kochhar, J. Leathersich, C. Moe, M. Winters, J. Shealy, and R. H. Olsson, "A Manufacturable AlScN Periodically Polarized Piezoelectric Film Bulk Acoustic Wave Resonator (AlScN P3F BAW) Operating in Overtone Mode at X and Ku Band," in *2023 IEEE/MTTS International Microwave Symposium - IMS 2023*, pp. 891–894, June 2023. ISSN: 2576-7216.

[3] M. Feng, S.-C. Shen, D. Caruth, and J.-J. Huang, "Device technologies for RF front-end circuits in next-generation wireless communications," *Proceedings of the IEEE*, vol. 92, pp. 354–375, Feb. 2004.

[4] P. Tedeschi, S. Sciancalepore, and R. Di Pietro, "Satellite-based communications security: A survey of threats, solutions, and research challenges," *Computer Networks*, vol. 216, p. 109246, 2022.

[5] G. Gultepe, T. Kanar, S. Zihir, and G. M. Rebeiz, "A 1024-Element Ku-Band SATCOM Dual-Polarized Receiver With >10-dB/K G/T and Embedded Transmit Rejection Filter," *IEEE Transactions on Microwave Theory and Techniques*, vol. 69, pp. 3484–3495, July 2021.

[6] M. Akiyama, T. Kamohara, K. Kano, A. Teshigahara, Y. Takeuchi, and N. Kawahara, "Enhancement of piezoelectric response in scandium aluminum nitride alloy thin films prepared by dual reactive cosputtering," *Advanced Materials*, vol. 21, no. 5, pp. 593–596, 2009.

[7] R. H. Olsson, Z. Tang, and M. D'Agati, "Doping of aluminum nitride and the impact on thin film piezoelectric and ferroelectric device performance," in *2020 IEEE Custom Integrated Circuits Conference (CICC)*, pp. 1–6, 2020.

[8] L. Chen, C. Liu, M. Li, W. Song, W. Wang, Z. Wang, N. Wang, and Y. Zhu, "Scandium-Doped Aluminum Nitride for Acoustic Wave Resonators, Filters, and Ferroelectric Memory Applications," *ACS Applied Electronic Materials*, vol. 5, pp. 612–622, Feb. 2023. Publisher: American Chemical Society.

[9] K. Saha, P. Simeoni, L. Colombo, and M. Rinaldi, "Piezoelectric and ferroelectric measurements on casted target-deposited Al0.45Sc0.45B0.1N thin films," *Frontiers in Materials*, vol. 12, p. 1567614, Apr. 2025. Publisher: Frontiers.

[10] R. Vetury, A. Kochhar, J. Leathersich, C. Moe, M. Winters, J. Shealy, and R. H. Olsson, "A Manufacturable AlScN Periodically Polarized Piezoelectric Film Bulk Acoustic Wave Resonator (AlScN P3F BAW) Operating in Overtone Mode at X and Ku Band," in *2023 IEEE/MTT-S International Microwave Symposium - IMS 2023*, pp. 891–894, 2023.

[11] L. Colombo, L. Spagnuolo, K. Saha, G. Giribaldi, P. Simeoni, and M. Rinaldi, "Scaln-on-sic k-band sezawa solidly-mounted bidimensional mode resonators," *IEEE Electron Device Letters*, vol. 46, no. 4, pp. 660–663, 2025.

[12] O. Barrera, N. Ravi, K. Saha, S. Dasgupta, J. Campbell, J. Kramer, E. Kwon, T.-H. Hsu, S. Cho, I. Anderson, P. Simeoni, J. Hou, M. Rinaldi, M. S. Goorsky, and R. Lu, "18 GHz Solidly Mounted Resonator in Scandium Aluminum Nitride on SiO/Ta O Bragg Reflector," *Journal of Microelectromechanical Systems*, pp. 1–0, 2024. 10.1109/JMEMS.2024.3472615.

[13] C. Cassella, G. Chen, Z. Qian, G. Hummel, and M. Rinaldi, "Crosssectional lame mode ladder filters for uhf wideband applications," *IEEE Electron Device Letters*, vol. 37, no. 5, pp. 681–683, 2016.

[14] G. Giribaldi, L. Colombo, P. Simeoni, and M. Rinaldi, "Compact and wideband nanoacoustic pass-band filters for future 5G and 6G cellular radios," *Nature Communications*, vol. 15, p. 304, Jan. 2024.

[15] L. Colombo, G. Giribaldi, and M. Rinaldi, "18 GHz Microacoustic ScAlN Lamb Wave Resonators for Ku Band Applications," in *2024 IEEE International Microwave Filter Workshop (IMFW)*, pp. 84–86, Feb. 2024.

[16] L. Colombo, G. Giribaldi, R. Tetro, J. Guida, W. Gubinelli, L. Spagnuolo, N. Maietta, S. Ghosh, and M. Rinaldi, "Characterization of Acoustic Losses in Interdigitated VHF to mmWave Piezoelectric M/NEMS Resonators," in *2024 IEEE Ultrasonics, Ferroelectrics, and Frequency Control Joint Symposium (UFFC-JS)*, pp. 1–4, Sept. 2024. ISSN: 23750448.

[17] J. Black, "Electromigration—A brief survey and some recent results," *IEEE Transactions on Electron Devices*, vol. 16, pp. 338–347, Apr. 1969.

[18] M. Braunovic, N. K. Myshkin, and V. V. Konchits, *Electrical Contacts Fundamentals, Applications and Technology*. 1st edition ed., 2007.

[19] J. Larson, P. Bradley, S. Wartenberg, and R. Ruby, "Modified Butterworth-Van Dyke circuit for FBAR resonators and automated measurement system," in *2000 IEEE Ultrasonics Symposium. Proceedings. An International Symposium (Cat. No.00CH37121)*, vol. 1, pp. 863–868 vol.1, Oct. 2000. ISSN: 1051-0117.

[20] G. Giribaldi, L. Colombo, and M. Rinaldi, "6–20 GHz 30% ScAlN Lateral Field-Excited Cross-Sectional Lamé Mode Resonators for Future Mobile RF Front Ends," *IEEE Transactions on Ultrasonics, Ferroelectrics, and Frequency Control*, vol. 70, pp. 1201–1212, Oct. 2023.

[21] S. Cho, O. Barrera, P. Simeoni, E. N. Marshall, J. Kramer, K. Motoki, T.H. Hsu, V. Chulukhadze, M. Rinaldi, W. A. Doolittle, and R. Lu, "Millimeter Wave Thin-Film Bulk Acoustic Resonator in Sputtered Scandium Aluminum Nitride," *Journal of Microelectromechanical Systems*, vol. 32, pp. 529–532, Dec. 2023.

[22] R. Tetro, L. Colombo, W. Gubinelli, G. Giribaldi, and M. Rinaldi, "XCut Lithium Niobate S0 Mode Resonators for 5G Applications," in *2024 IEEE 37th International Conference on Micro Electro Mechanical Systems (MEMS)*, pp. 1102–1105, Jan. 2024.

[23] G. Giribaldi, L. Colombo, M. Castellani, M. A. Masud, G. Piazza, and M. Rinaldi, "Up-Scaling Microacoustics: 20 to 35 GHz ALN Resonators with f · Q Products Exceeding 14 THz," in *2024 IEEE 37th International Conference on Micro Electro Mechanical Systems (MEMS)*, pp. 31–35, Jan. 2024.

[24] W. Zhao, M. J. Asadi, L. Li, R. Chaudhuri, K. Nomoto, H. G. Xing, J. C. M. Hwang, and D. Jena, "15-GHz Epitaxial AlN FBARs on SiC Substrates," *IEEE Electron Device Letters*, vol. 44, pp. 903–906, June 2023.